\documentclass[ 
 superscriptaddress,
nofootinbib,
 amsmath,amssymb,
 aps,
 twocolumn,
 pra,
]{revtex4-2}

\usepackage{xcolor}
\usepackage{comment}
\usepackage{graphicx}
\usepackage{dcolumn}
\usepackage{bm}
\usepackage[colorlinks=true,linkcolor=blue,citecolor=blue]{hyperref}
\usepackage{cancel}
\usepackage{tikz}

\def\selectlanguage#1{}

\begin{document}

\preprint{APS/123-QED}

\title{Experimental Demonstration of 7-Dimensional N-Vector Model with\\ Dimensional Inflation and Annealing}
\author{Daniele Veraldi}
\email{daniele.veraldi@uniroma1.it}
\affiliation{Department of Physics, Sapienza University, 00185 Rome, Italy}
\author{Marcello Calvanese Strinati}
\affiliation{Enrico Fermi Research Center (CREF), Via Panisperna 89a, 00184 Rome, Italy}
\author{Paolo Charlesworth}
\affiliation{Department of Physics, Sapienza University, 00185 Rome, Italy}
\author{Matteo Spaziani}
\affiliation{Department of Physics, Sapienza University, 00185 Rome, Italy}
\author{Silvia Gentilini}
\affiliation{Institute for Complex Systems, National Research Council, 00185 Rome, Italy}
\author{Davide Pierangeli}
\affiliation{Department of Physics, Sapienza University, 00185 Rome, Italy}
\author{Claudio Conti}
\affiliation{Department of Physics, Sapienza University, 00185 Rome, Italy}
\affiliation{Enrico Fermi Research Center (CREF), Via Panisperna 89a, 00184 Rome, Italy}
\date{\today}
\begin{abstract}
  We report the first experimental realization of a physical system described by a high-dimensional
  Hamiltonian comprising up to seven dimensions.
  Our setup is based on a holographic optical computer, which realizes a dimensional inflation by
  spatial and temporal multiplexing of lower dimensional systems.
  We demonstrate an $N$-vector model, with applications in field theory and computing.
  The approach is scalable and enables table-top implementation of high-dimensional theoretical models,
  and novel methods for combinatorial optimization.
  We experimentally demonstrate dimensional annealing: by dynamically reducing the number of dimensions, the ground state of an Ising model is found in complex graphs by an adiabatic evolution from high-dimensional spaces.
\end{abstract}
\maketitle

Whether in fundamental physics or in large language models, the dimensionality of the representation plays a crucial role.
Since Theodor Kaluza~\cite{kaluza_unification_2018}, physicists adopted artificial dimensions for the unification of forces, such as gravity and electromagnetism. String theory further exposed the path towards grand-unification by eleven or more spatial dimensions~\cite{witten_string_1995,polchinski_string_1998,polyakov_gauge_2018}.
Recently, the discovery of the duality between strings and the $N$-vector model has broadened the landscape~\cite{klebanov_ads_2002}. The $N$-vector model ($N$ is the number of dimensions) is a paradigm for universal classes~\cite{guida_critical_1998}:
it represents Ising spins for $N=1$, superfluid helium for $N=2$, ferromagnets for $N=3$, and
the Higgs sector of the standard model ($N=4$), with a role in the thermal transition of quarks~\cite{aarts_phase_2023}.
In statistical mechanics~\cite{stanley_dependence_1968}, $N$-vector model is a cornerstone for the understanding of critical phenomena~\cite{wilson_critical_1972,parisi_statistical_1988}.
However, beyond theoretical models, no experiment helped to shade light on the open issues with high-dimensionality, such as the effects of boundaries~\cite{metlitski_boundary_2021} or the interpolation between small- and large-$N$ regimes~\cite{kos_bootstrapping_2014}. We do not have experimental devices for multidimensional physics at a large $N$, or when $N$ varies with evolution.

On a different perspective, algorithms for artificial intelligence make high-$N$ models compelling. Indeed, non-separable datasets can be classified when represented in high-dimensional feature spaces~\cite{scholkopf_learning_2018}. Multidimensional embedding is at the inner core of large language models as it captures semantic, syntactic and contextual relationships between word tokens through the attention mechanism~\cite{mikolov_efficient_2013,vaswani_attention_2017}. Recently, authors have shown that high-dimensional vector models accelerate combinatorial optimization~\cite{PhysRevLett.126.143901,calvanese_strinati_multidimensional_2022,Ben-Ami:23,strinati_optical_2026,cummins_vector_2025,syed_soft_2026}. New strategies, such as dimensional annealing, where the optimization
initially occurs at large-$N$ to progressively project to the target dimensionality, increases by order of magnitudes the probability of success~\cite{calvanese_strinati_hyperscaling_2024,calvanese_strinati_equalized_2025,chiavazzo_ising_2025}.

Demonstrating physical devices that can explore spaces with large $N$ is significant for fundamental physics and for modern algorithms. We report the first experiments on the $N$-vector model, and the first experimental realization of dimensional annealing. We exploit a massively parallel optical computer by introducing two innovations: First, a mathematical formulation that composes a high-dimensional Hamiltonian from a number of coupled low-dimensional systems, a process that we name dimensional inflation. Second, a photonic device that exploits spatio-temporal multiplexing to compute the low dimensional Hamiltonians and combine them to achieve large $N$ regimes. The high-dimensional Hamiltonian is determined with fidelity by the large-scale and inherent parallelism of holography, which also enables to change $N$ dynamically to demonstrate dimensional annealing.

To illustrate the idea of dimensional inflation, we consider the Heisenberg Hamiltonian $H_N$ with $N=3$ of $L$ coupled spins $\mathbf{S}_j$, represented by three-dimensional vectors with unitary length. The Hamiltonian in spherical coordinates is
\begin{eqnarray}
    H_{3}&=&-\sum_{j,k=1}^{L}J_{jk}\mathbf{S}_{j}\cdot\mathbf{S}_{k}\nonumber\\
          &=&-J \sum_{j,k=1}^{L}\sin(\theta_{j})\sin(\theta_{k})\cos(\varphi_{j}-\varphi_{k})-J \sum_{j,k=1}^{L}\sigma_{j}\sigma_{k}\nonumber\\
          &=&H_{2}+H_{1}\,\, ,
\label{eq:2}\end{eqnarray}
where $\sigma_{j}=\cos(\theta_{j})$.
Initially we consider $J_{jk}=J$. Equation~\eqref{eq:2} shows that the $H_{3}$ is actually the sum of the energy of two systems. One two-dimensional (``azimuthal'') Hamiltonian dependent on both the azimuthal and polar angles $\varphi_j$ and $\theta_j$, respectively, where $\sin(\theta_{j})\sin(\theta_{k})$ acts as an effective coupling coefficient. The second Hamiltonian describes the ``polar'' system that only depends on the polar angles $\theta_j$, and maps an Ising model $H_{1}=-J\sum_{j,k=1}^{L}\sigma_{j}\sigma_{k}$,
if we identify the (soft) Ising spins with $\sigma_{j}=\cos(\theta_{j})$. Notably, if the spins $\mathbf{S}_j$ belong to an $N$-dimensional space, the decomposition still holds with $N-2$ polar and one azimuthal systems (see Methods).
Namely, dimensional inflation can be formulated by the recursive relation:
\begin{equation}
H_{N-1}+H_{1}=H_{N}\;.
\label{eq:inflation}
\end{equation}
If we realize experimentally the $N-1$ polar-azymuthal systems, we can simulate an arbitrary dimensional Hamiltonian by their coupling. Also, if during the dynamics we reduce the Hamiltonian dependence on all angles to zero, except one polar angle $\theta_j$, we have a projection (i.e., the dimensional annealing) to an Ising model, which is pivotal for combinatorial optimization.
\begin{figure*}[htbp]
    \centering
    \includegraphics[width=1\linewidth]{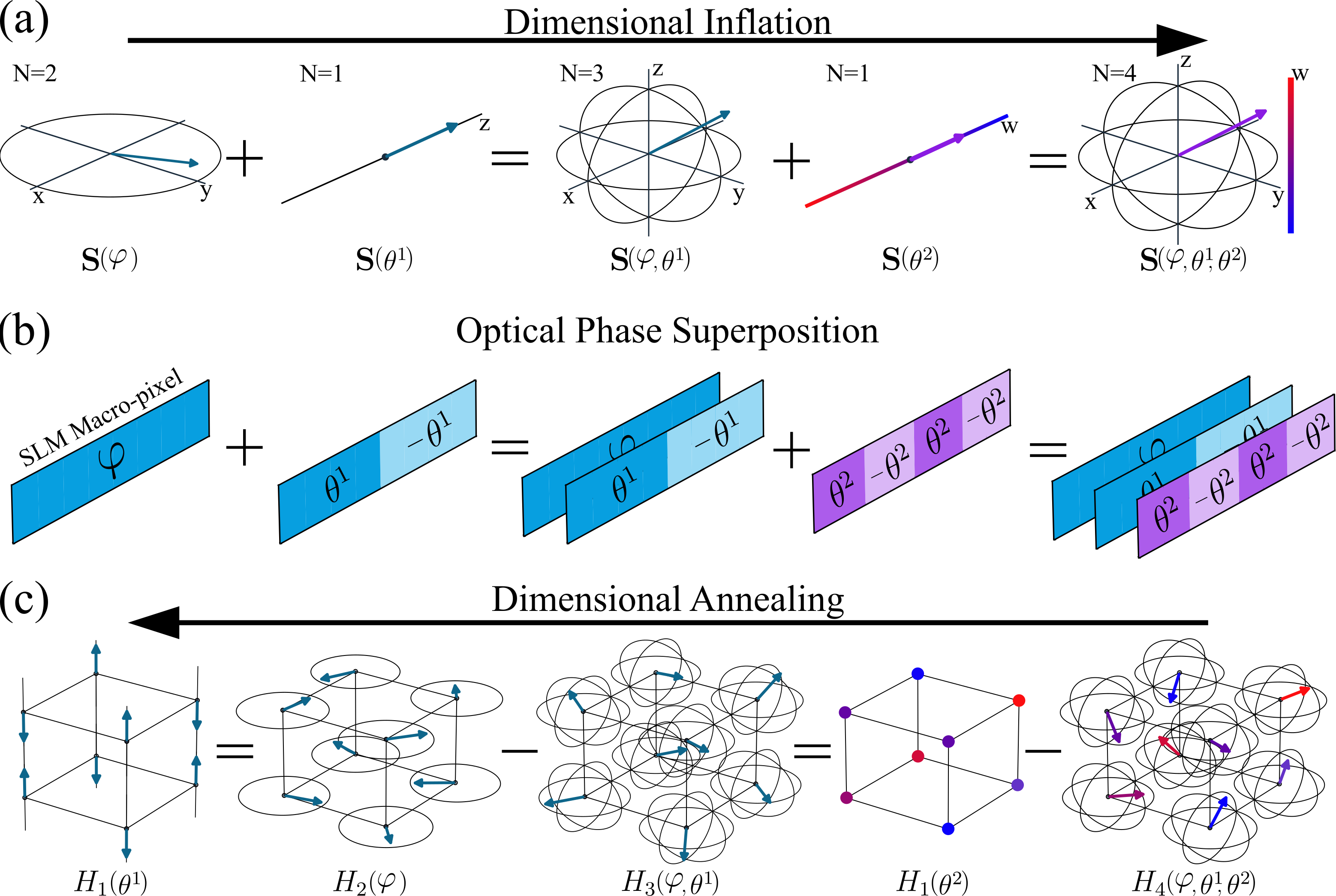}
    \caption{{\bf Dimensional Inflation and Annealing.} 
    (a) Dimensional inflation generates arbitrary $N$-vector spins by recursively composing low dimensional spins [Eq.~\eqref{eq:inflation}]. 
    (b) By superimposing multiple binary gratings and flat phase delay patterns on the SLM, we are able to optically encode multiple angles simultaneously.
    (c) Dimensional annealing process, which can be seen as the reverse process of dimensional inflation, allows reducing the system dimensionality by suppressing selected spin components.  }
    \label{fig:hamiltonian_composition}
\end{figure*}

In hyperspherical coordinates~\cite{kalnins2002ndimensional,cohl2011ndimensional},
a  unit vector $\mathbf{S}$ is parameterized by $N-2$ polar angles $\theta^p \in [0, \pi]$, with $p = 1, \dots, N-2$, and one azimuthal angle $\varphi \in [0, 2\pi]$:

\begin{equation}
  \mathbf{S}
\!=\!
\begin{pmatrix}
    \cos(\theta^{1})\\
\vdots \\
\left[ \prod_{p=1}^{l-1} \sin(\theta^{p}) \right] \cos(\theta^{l}) \\
\vdots \\
\left[ \prod_{p=1}^{N-2} \sin(\theta^{p}) \right] \cos(\varphi)\\
\left[ \prod_{p=1}^{N-2} \sin(\theta^{p}) \right] \sin(\varphi)
\end{pmatrix} \,,
\end{equation}
where $l=1\,\ldots,N-2$, with $l=1$ being the first component $\cos(\theta^1)$. As detailed in methods, the $N$-dimensional scalar product, $\mathbf{S}_j \cdot \mathbf{S}_k$, is a linear combination of scalar products components in lower dimensions. Therefore, a high dimensional Hamiltonian is a combination of lower dimensional Hamiltonians (see Fig.~\ref{fig:hamiltonian_composition}).

Such a decomposition also holds when considering the arbitrary couplings $J_{jk}$, by considering it as a further artificial dimension.
Indeed, for $L$ interacting spins, the Hamiltonian $H_N$ is diagonalized into a sum of $K$, rank-1 (Mattis) Hamiltonians~\cite{yamashita_low-rank_2023}, where $K$ is the rank of the $L\times L$ interaction matrix $J_{jk}$:
\begin{eqnarray}
    H_N &=& -\sum_{j < k} J_{jk} \, \mathbf{S}_j \cdot \mathbf{S}_k \nonumber\\
    &=& -\sum_{m=1}^K  \lambda_m \sum_{j < k} \xi_{j,m} \xi_{k,m} \, \mathbf{S}_{j} \cdot \mathbf{S}_{k} \, ,
    \label{eq:rankN_decomposition}
\end{eqnarray}
where $m \in \{1, \dots, K\}$ indexes the modes and $\lambda_m$ represents the eigenvalue weight of the $m$-th mode
The interaction coefficient $\xi_j \in [-1, 1]$ can be interpreted as an additional polar angle $\theta^{J}_j= \arccos(\xi_j) \in [0, \pi]$. Thus, we define an interacting spin by the set of $N$ angles $\{\varphi, \theta^{J}, \theta^{1}, \dots, \theta^{N-2} \}$.

\begin{figure*}[htbp]
    \centering
    \includegraphics[width = \textwidth]{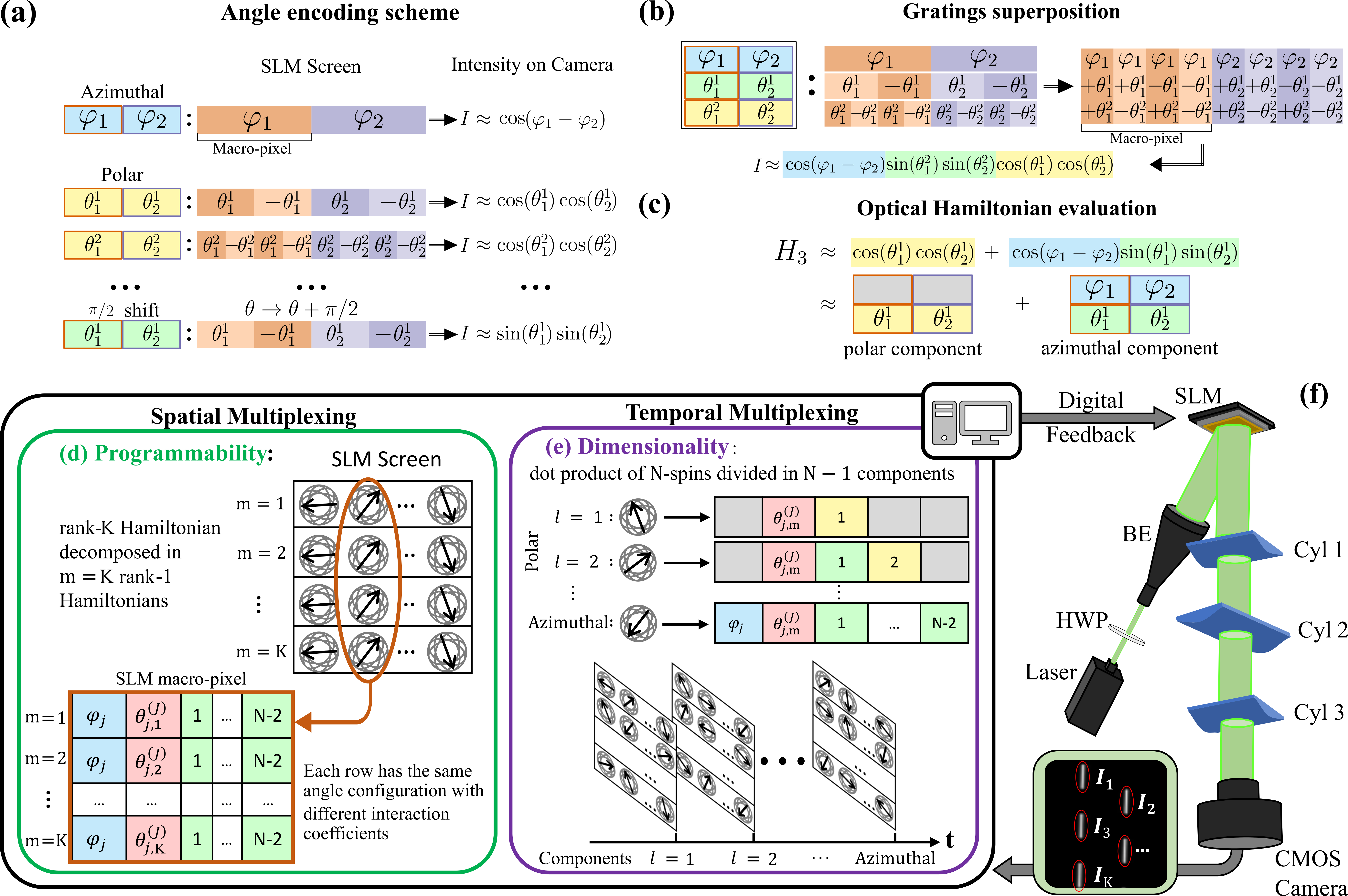}
    \caption{{\bf Optical implementation by spatio-temporal multiplexed phases.} 
    (a) Angles encoding scheme. For each macro-pixel $j$, the azimuthal angle $\varphi_j$ is encoded as a uniform phase delay across the macro-pixel, while the polar angles $\theta^{p}_j$ are encoded using binary phase gratings with periods $T_p=2^{p}$. By looking at the intensity contributions on the camera plane, we can evaluate Eq.~\eqref{eq:intensity_cos_delta_phi} and Eq.~\eqref{eq:intensity_gauge}. 
    (b) The superposition of phase gratings allows encoding multiple angles on the same macropixel, optically evaluating Eq.~\eqref{eq:tot_intensity_mod}.
    (c) By combining multiple optical evaluation, one for each scalar product component, we are able to evaluate the $N$-dimensional spin-spin interaction term.       
    (d) Spatial multiplexing for matrix programmability: A rank-$K$ interaction Hamiltonian is spatially multiplexed in $K$ rank-1 Mattis Hamiltonians across $m = 1, \dots, K$ macro-pixel rows on a phase-only SLM. Each row encodes identical $N$-vector spin configurations modulated by distinct coupling coefficients $\theta_{j,m}^{(J)}$, projecting $K$ spatially resolved intensity signals ($I_1, \dots, I_K$) onto the camera plane. 
    (e) Temporal multiplexing for spin dimensionality: $N$-dimensional scalar products are evaluated across $N-1$ consecutive temporal shots ($l$). Phase profiles corresponding to $N-2$ polar components ($l = 1, \dots, N-2$) and the azimuthal component are loaded sequentially onto the SLM. 
    (f) Experimental optical setup.
    }
    \label{fig:exp_setup}
\end{figure*}
The previous arguments show that to realize high-dimensional spins with arbitrary interactions we need a device that can store information of $N$ angles; we adopt a free-space spatial optical computing platform with a spatial light modulator (SLM).
Our architecture combines focal plane division (FPD) spatial multiplexing for full coupling matrix programmability with temporal multiplexing over $N-1$ consecutive shots to optically evaluate $N$-dimensional vector Hamiltonians~\cite{veraldi_fully_2025}.
The SLM encodes the information regarding the spin configuration and the interaction matrix into the phase profile of a laser beam,~\cite{veraldi_fully_2025} while the spin-spin interaction energy is computed by propagating the light through a lens and then detected as an intensity pattern on a camera (Fig.~\ref{fig:exp_setup}).
The key idea is to assign to each $N$-spin a subregion of the SLM of $P_x\times P_y$ pixels, called macro-pixel, and encode in each macro-pixel a set of angles for each scalar product to evaluate.
Starting from the azimuthal angle $\varphi$, we encode a uniform phase delay $e^{i\varphi}$ to the entire macro-pixel. As shown in Fig.~\ref{fig:exp_setup}(a), after propagating the light through the optical system, the interference between two macro-pixels $j$ and $k$ produce an intensity modulation:
\begin{equation}
    I \propto \cos(\varphi_j - \varphi_k)\,\, ,
    \label{eq:intensity_cos_delta_phi}
\end{equation}
at the origin of the Fourier plane. 
The polar angles $\{\theta^{J},\theta^{1}, \dots, \theta^{N-2}\}$ are encoded by using binary phase gratings~\cite{mendoza-yero_encoding_2014, luo_wavelength-division_2023} with phase values $\pm\theta$.
The interference between two binary gratings for the angles $\theta_j$ and $\theta_k$ produces an intensity modulation:
\begin{equation}
I \propto \cos(\theta_j)\cos(\theta_k) \,\, ,
\label{eq:intensity_gauge}
\end{equation} at the origin of the Fourier plane (Fig.~\ref{fig:exp_setup}(a)).

To encode multiple angles within a single macro-pixel, we assign each angle to a different binary phase grating, each characterized by distinct spatial periods $T_p= 2^{p}$, where $p=1,\ldots,N-2$. The gratings are superimposed on the SLM, creating a complex multi-level phase modulation whose amplitude transmittance is the product of the individual transmittances (Fig.~\ref{fig:exp_setup}(b)).
Consequently, when two macro-pixels $j$ and $k$ interfere, the total intensity modulation at the origin of the Fourier plane is given by:
\begin{equation}
    I \propto \prod_p \cos(\theta^{p}_j) \cos(\theta^{p}_k) \cos(\phi_j-\phi_k) 
    \,\, .
    \label{eq:tot_intensity_mod}
\end{equation}

As shown in Fig.~\ref{fig:exp_setup}(c), by combining Eq.~\eqref{eq:intensity_cos_delta_phi} with Eq.~\eqref{eq:intensity_gauge} we optically compute any scalar product component between spin $j$ and $k$ that appear in $H_{N}$. The sum over all pairs of macro-pixels is obtained by the interference on the camera plane in a single shot.

Finally, to implement arbitrary interactions, as in Eq.~\eqref{eq:rankN_decomposition}, we need to perform $K$ optical evaluations. Overall, $K(N-1)$ independent optical evaluations are required to compute an $N$-dimensional Hamiltonian with arbitrary couplings. To evaluate the $K(N-1)$ required components, we implement spatio-temporal multiplexing (Fig.~\ref{fig:exp_setup}(d,e)): the FPD processes the scalar product component of $K$ Mattis Hamiltonians simultaneously across SLM macro-pixel rows, while temporal multiplexing evaluates the $N-1$ scalar product components sequentially over successive frames.
This flexible spatio-temporal trade-off enables optimal resource utilization based on the total spin count $L$, vector dimension $N$, and macro-pixel sizes $P_x$ and $P_y$.

The experimental setup is shown in Fig.~\ref{fig:exp_setup}(f). A continuous-wave laser beam ($\lambda = 532\text{~nm}$) passes through a half-wave plate (HWP) and beam expander (BE) to illuminate a phase-only spatial light modulator (Hamamatsu LCOS-SLM X15213S, $1280 \times 1024$ pixels). Phase profiles are mapped using 213 calibrated gray levels over the $[0, 2\pi]$ phase range. An anamorphic lens system consisting of three cylindrical lenses ($\text{Cyl}_1$ with $f_1 = 250\text{~mm}$, $\text{Cyl}_2$ with $f_2 = 400\text{~mm}$, and $\text{Cyl}_3$ with $f_3 = 150\text{~mm}$) focuses the modulated field along the $x$-axis ($\text{Cyl}_2$), while images it along the $y$-axis ($\text{Cyl}_1$ and $\text{Cyl}_3$) onto a 16-bit Complementary Metal-Oxide-Semiconductor (CMOS) camera (Hamamatsu ORCA-Fusion C14440-20UP). This configuration generates $K$ spatially separated intensity spots on the sensor plane, corresponding to the simultaneous energy evaluations of $K$ Mattis components. The system runs under a feedback loop controlled by a digital processor.

\begin{figure*}[htbp] 
    \centering
    \includegraphics[width=\linewidth]{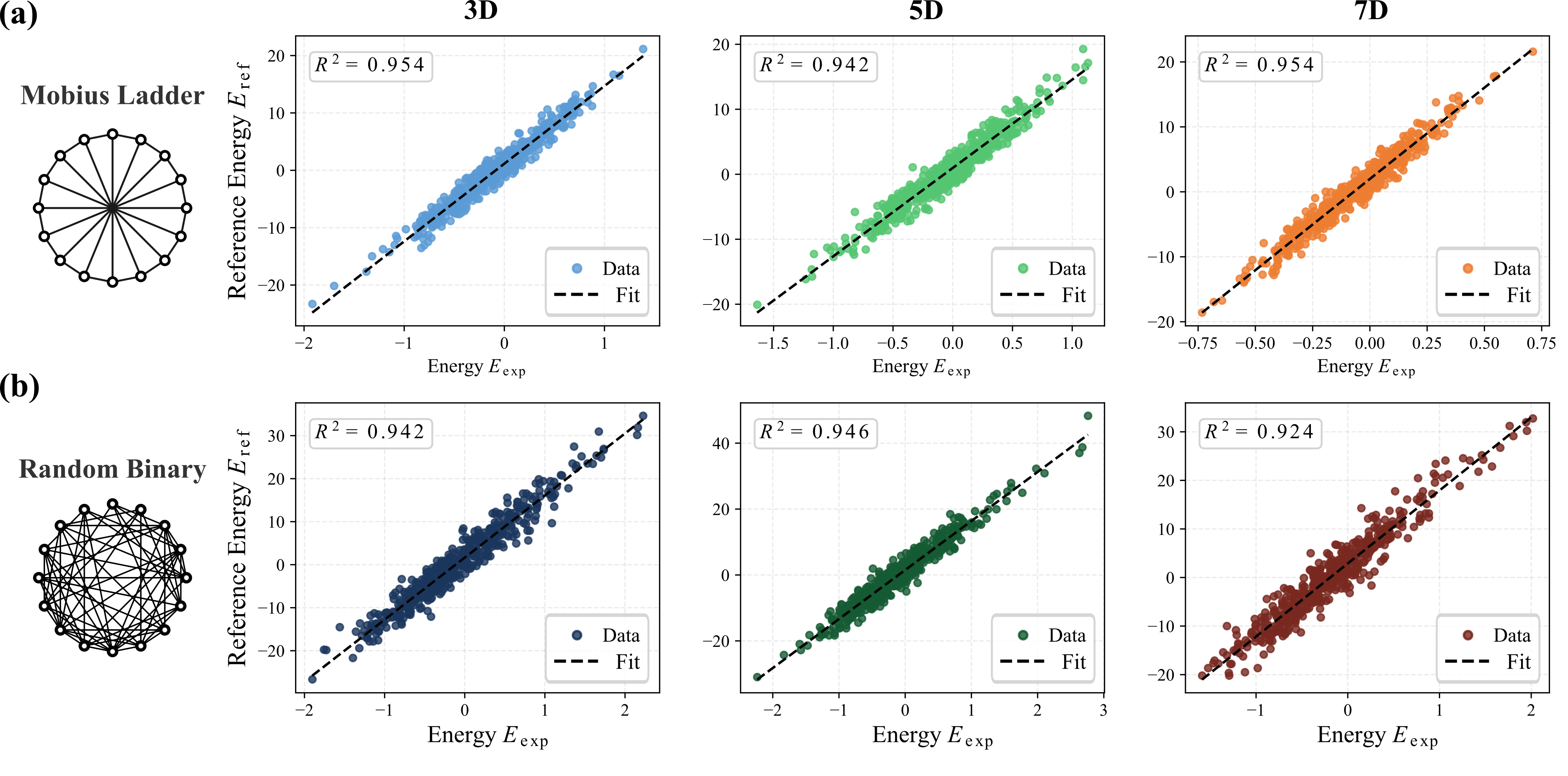}
    \caption{{\bf Fidelity of optical Hamiltonian evaluation for 16-node graphs.} 
    We compare experimental ($E_{\text{exp}}$) and theoretical reference ($E_{\text{ref}}$) values across $500$ random $N$-spin states for (a) Möbius ladder and (b) random binary graphs. Results showcase the agreement between experimental and theoretical values for spins of dimensions $N=3,5,7$. All the experiments are performed under the same experimental conditions: $L=16$ spins, $32\times64$ macro-pixel footprint and same graph structure.}
    \label{fig:calibrations}
\end{figure*}

We experimentally simulate the high-dimensional Hamiltonian in Eq.~\ref{eq:rankN_decomposition}, successfully encoding and evaluating Hamiltonians of $L=16$ spins and up to $N=7$ dimensions. We evaluate the fidelity of the simulation by comparing the experimentally measured Hamiltonian with its analytical counterpart for $500$ random configurations of $N$-vector spins for $N=3,5,7$. In particular, we consider two types of graphs that are widely used as benchmarks for Ising Machines (IMs) \cite{mohseni_ising_2022}: M\"obius ladder and random binary graphs. M\"obius ladder is a circulant graph~\cite{davis1994circulant} where every spin interacts antiferromagnetically $J_{ij} = -1$ with its nearest neighbors and the diametrically opposite spin, while the binary graph is randomly generated by extracting spin couplings  $J_{ij}=0$ or $J_{ij}=-1$ with $0.5$ probability. To have a fair comparison of the performances at increasing $N$, we fix the total number of spins to $L=16$ and the macro-pixel size to $32\times64$ pixels. The results are shown in Fig.~\ref{fig:calibrations}. To have a quantifiable metric fidelity, we use the coefficient of determination $R^2$. We observe that the coefficient of determination is $R^2\approx0.95$ for both graphs and doesn't show significant trend when scaling to higher spin dimension, ensuring that the performances are not affected when increasing the number of dimensions of the spins. Residual scatter around the ideal unity response ($E_{\text{exp}} = E_{\text{ref}}$) is governed by deterministic wavefront phase aberrations, non-uniform spatial illumination intensity across the SLM aperture, 8-bit SLM phase quantization steps and CMOS camera shot noise.

Furthermore, we report the first experimental implementation of the dimensional annealing algorithm.
First introduced in the context of the hyperspin machine~\cite{calvanese_strinati_multidimensional_2022,calvanese_strinati_hyperscaling_2024}, dimensional annealing is an optimization heuristic designed to escape local energy traps by tuning during the dynamics the hyperspin dimensionality, from $N'$ to $N$. When $N<N'$, the system initially relaxes towards a low-energy state within the higher, $N'$-dimensional space, where local minima of the target $N$-vector model are escaped thanks to the higher vector dimension, before gradually reducing the dimensionality in time and converge to low-energy states of the target Hamiltonian in dimension $N$~\cite{calvanese_strinati_multidimensional_2022,calvanese_strinati_hyperscaling_2024}.

\begin{figure*}[htbp] 
    \centering
    \includegraphics[width=\linewidth]{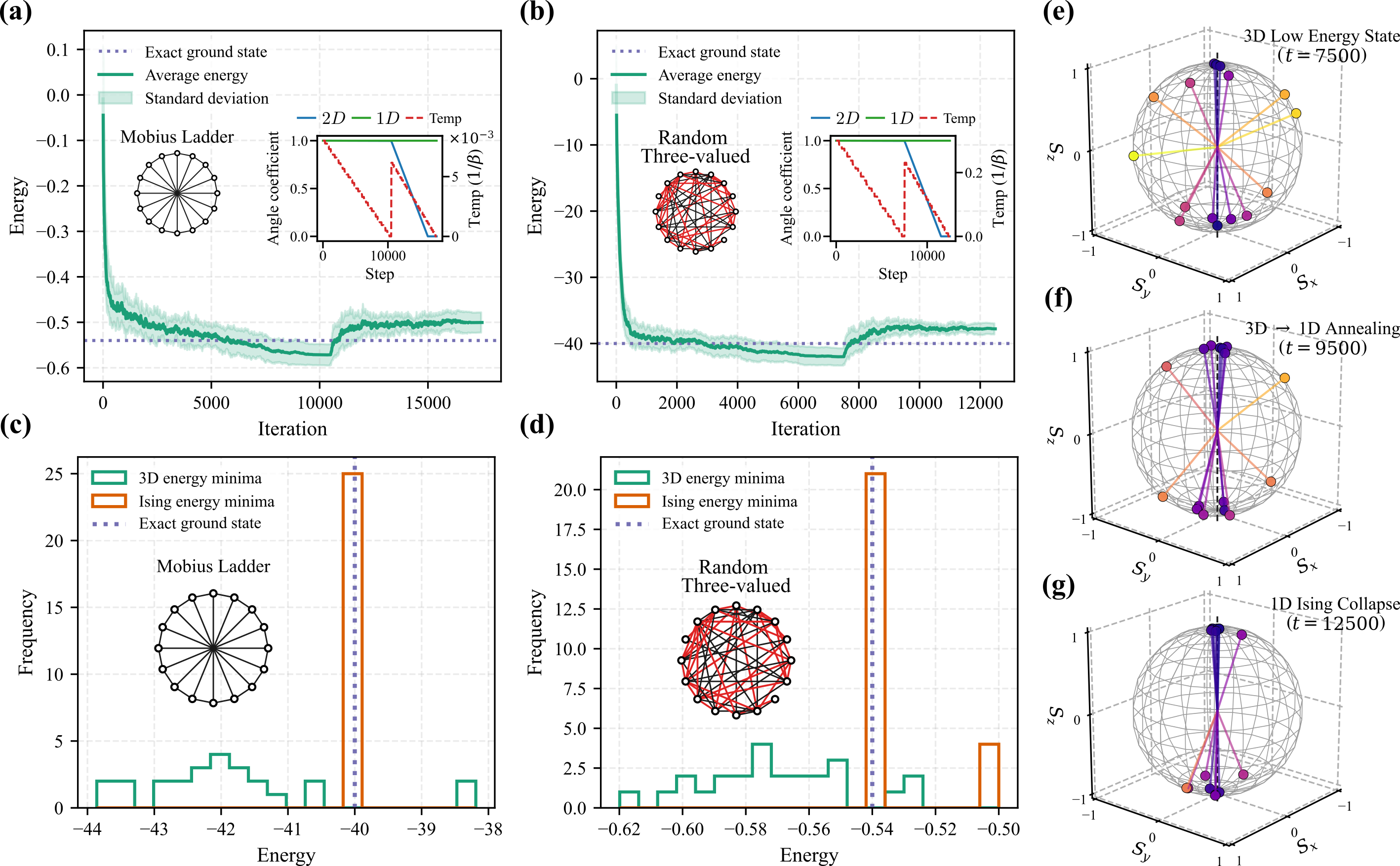}
    \caption{{\bf Dimensional annealing dynamics and energy distribution for 16-node graphs.}
    (a) Energy evolution (mean: solid line; standard deviation across 25 runs: shaded region) during $N=3 \to N=1$ dimensional annealing on a M\"{o}bius ladder graph, along with the ground state energy of the Ising model (purple dotted line). Inset: control schedule detailing the temperature $1/\beta$ (red dashed line) and angular coupling coefficients for the one-dimensional (green) and two-dimensional (blue) subspace components. (b) Corresponding energy trajectory and parameter schedule for a 16-node random three-valued graph under the same annealing protocol. (c, d) Energy histograms over 25 independent runs for (c) M\"{o}bius ladder and (d) random three-valued graphs, comparing continuous $N=3$ energy minima (green) and post-annealing binarized Ising energies (orange) against exact ground state baselines (purple dotted lines). (e, f, g) Spins configuration for the M\"{o}bius ladder graph at iterations (e) $t=7500$, (f) $t=9500$, and (g) $t=12500$, depicting the collapse of continuous three-dimensional spins into discrete polar Ising states ($S_z = \pm 1$). 
    }
    \label{fig:dim-ann-distribution}
\end{figure*}

In our experimental simulation of the $N$-vector model, the dimensional annealing is performed by assigning a coefficient to each component of the scalar product and then slowly bringing to zero the coefficients relative to the components of the scalar product that contains the information about the dimensions that we are annealing. For instance, if we look at the $N=3 \rightarrow N=1$ case (i.e., from Heisenberg to Ising spins defined along the $z$-axis), we see that the one-dimensional vector component $z$ (see Fig.~\ref{fig:exp_setup}(c)) depends only on the polar angle $\theta$ of the three-dimensional vector spins, while the remaining two-dimensional component on the transverse $xy$-plane depends on both the polar $\theta$ and the azimuthal angle $\varphi$. Therefore, by slowly decreasing the coefficient of the two-dimensional component while keeping the coefficient of the one-dimensional component constant we let the system collapse into the lower-dimensional subspace, effectively driving the system from a low-energy state in three dimensions to a low-energy state in one dimension, in a controlled way, bypassing the local minima of the target Ising energy.

The dimensional annealing protocol guides the system transition between different subspaces, while a simulated annealing algorithm explores the configuration space to find the lowest energy states.  

Figure~\ref{fig:dim-ann-distribution}(a-b) shows the average energy trajectory across 25 independent runs of dimensional annealing for two 16-node graphs: a M\"{o}bius ladder graph and a random three-valued graph~\cite{calvanese_strinati_equalized_2025} with edge density $\rho = 0.6$. The inset of each panel displays the temperature schedule used by the simulated annealing and the coefficient schedule for the dimensional annealing stage. In the first phase, the temperature drops linearly to drive the system to a low-energy state of the $N=3$ Hamiltonian. Next, the dimensional annealing schedule begins: the coefficient governing the two-dimensional components decreases linearly to zero, while the temperature undergoes a step-up and subsequent linear decay to extract the system from local minima in $N=3$ towards the ground state of the $N=1$ target Hamiltonian. Notably, the energy at the end of the algorithm remains higher than the discrete Ising ground-state energy. This behavior is expected because the optimization outputs continuous (soft) one-dimensional spin configurations that must undergo post-processing binarization. As demonstrated in Fig.~\ref{fig:dim-ann-distribution}(c, d), the resulting binarized spin configurations correspond to the ground states of the target Ising Hamiltonians, achieving success accuracies of $100\%$ and $84\%$ for the M\"{o}bius ladder and random three-valued graphs, respectively.
Figure~\ref{fig:dim-ann-distribution}(e) illustrates the collapse of the $N=3$ spins during dimensional annealing. Examining the axial ($S_z$) and transverse ($S_\perp$) spin components reveals that, at the beginning of the dimensional annealing, the spin vectors are distributed across the entire spherical configuration space. As the annealing process proceeds, the spins migrate toward the two polar regions as their transverse components decay ($S_\perp \to 0$). At the end of the dimensional annealing phase, the spins cluster tightly at the north or south pole ($S_z = \pm 1$). From this state, discrete binary spin values are assigned based on proximity to the nearest pole.

In this work, we have reported the first experimental simulation of a seven-dimensional physical system by dimensional inflation and dimensional annealing protocols. A new holographic approach that extends spatial optical computing beyond binary discrete spin variables ($S_j \in \{-1, +1\}$) to continuous $N$-dimensional vector spin models ($\mathbf{S}_j \in \mathbb{R}^N$). By using a hybrid spatio-temporal multiplexing strategy, paired with frequency-spaced binary gratings encoding, our platform optically evaluates $N$-vector spin Hamiltonians while preserving full coupling matrix programmability through low-rank Mattis decomposition.

These achievements open several promising avenues for future research in optical hardware development, fundamental physics validations, and hybrid computing. While the temporal multiplexing scheme with $K(N-1)$ frames in the current setup is rate-limited by the liquid-crystal SLM refresh rate ($\sim 60$--$120$~Hz), replacing liquid-crystal modulators with high-speed ferroelectric SLMs or integrated Micro-Electro-Mechanics-Systems (MEMS) spatial light modulators operating at kilohertz-to-megahertz frequencies will boost optical evaluation throughput by several orders of magnitude.

Beyond computational hardware acceleration, the capability to continuously map and control $N$-dimensional vector spins provides a versatile physical testing-ground for fundamental physics. The platform can be directly applied to simulate complex statistical mechanics phase transitions, such as Kosterlitz-Thouless transitions in XY models ($N=2$), continuous Heisenberg spin glasses ($N=3$), renormalization flows with varying dimension, and topological defect dynamics in high-dimensional non-Abelian vector field theories. Finally, interfacing with closed-loop tensor processing units will establish hybrid optoelectronic computing architectures capable of solving large-scale continuous vector optimization problems, including phase retrieval, continuous vector packing, and non-convex energy landscape sampling.

\vspace*{0.1cm}
We acknowledge financial support from
HORIZON-ERC-2023-ADG HYPERSPIM ‘Optoelectronic and all-optical hyperspin machines for large-scale computing’ no. 101139828,
HORIZON-ERC-2024-STG LOOP ‘Optical polarization for ultrafast computing’ no. 101162173,
and HORIZON-EIC-2022 HEISINGBERG no. 101114978. Views and opinions expressed are however those of
the authors only and do not necessarily reflect those of the EU or the ERC. Neither the EU nor
the granting authority can be held responsible for them.
\bibliography{references_DV}

\begin{thebibliography}{32}%
\makeatletter
\providecommand \@ifxundefined [1]{%
 \@ifx{#1\undefined}
}%
\providecommand \@ifnum [1]{%
 \ifnum #1\expandafter \@firstoftwo
 \else \expandafter \@secondoftwo
 \fi
}%
\providecommand \@ifx [1]{%
 \ifx #1\expandafter \@firstoftwo
 \else \expandafter \@secondoftwo
 \fi
}%
\providecommand \natexlab [1]{#1}%
\providecommand \enquote  [1]{``#1''}%
\providecommand \bibnamefont  [1]{#1}%
\providecommand \bibfnamefont [1]{#1}%
\providecommand \citenamefont [1]{#1}%
\providecommand \href@noop [0]{\@secondoftwo}%
\providecommand \href [0]{\begingroup \@sanitize@url \@href}%
\providecommand \@href[1]{\@@startlink{#1}\@@href}%
\providecommand \@@href[1]{\endgroup#1\@@endlink}%
\providecommand \@sanitize@url [0]{\catcode `\\12\catcode `\$12\catcode
  `\&12\catcode `\#12\catcode `\^12\catcode `\_12\catcode `\%12\relax}%
\providecommand \@@startlink[1]{}%
\providecommand \@@endlink[0]{}%
\providecommand \url  [0]{\begingroup\@sanitize@url \@url }%
\providecommand \@url [1]{\endgroup\@href {#1}{\urlprefix }}%
\providecommand \urlprefix  [0]{URL }%
\providecommand \Eprint [0]{\href }%
\providecommand \doibase [0]{https://doi.org/}%
\providecommand \selectlanguage [0]{\@gobble}%
\providecommand \bibinfo  [0]{\@secondoftwo}%
\providecommand \bibfield  [0]{\@secondoftwo}%
\providecommand \translation [1]{[#1]}%
\providecommand \BibitemOpen [0]{}%
\providecommand \bibitemStop [0]{}%
\providecommand \bibitemNoStop [0]{.\EOS\space}%
\providecommand \EOS [0]{\spacefactor3000\relax}%
\providecommand \BibitemShut  [1]{\csname bibitem#1\endcsname}%
\let\auto@bib@innerbib\@empty
\bibitem [{\citenamefont {Kaluza}(2018)}]{kaluza_unification_2018}%
  \BibitemOpen
  \bibfield  {author} {\bibinfo {author} {\bibfnamefont {T.}~\bibnamefont
  {Kaluza}},\ }\href {https://doi.org/10.1142/S0218271818700017} {\bibfield
  {journal} {\bibinfo  {journal} {International Journal of Modern Physics D}\
  }\textbf {\bibinfo {volume} {27}},\ \bibinfo {pages} {1870001} (\bibinfo
  {year} {2018})},\ \bibinfo {note} {arXiv:1803.08616
  [physics.hist-ph]}\BibitemShut {NoStop}%
\bibitem [{\citenamefont {Witten}(1995)}]{witten_string_1995}%
  \BibitemOpen
  \bibfield  {author} {\bibinfo {author} {\bibfnamefont {E.}~\bibnamefont
  {Witten}},\ }\href {https://doi.org/10.1016/0550-3213(95)00158-O} {\bibfield
  {journal} {\bibinfo  {journal} {Nuclear Physics B}\ }\textbf {\bibinfo
  {volume} {443}},\ \bibinfo {pages} {85} (\bibinfo {year} {1995})}\BibitemShut
  {NoStop}%
\bibitem [{\citenamefont {Polchinski}(1998)}]{polchinski_string_1998}%
  \BibitemOpen
  \bibfield  {author} {\bibinfo {author} {\bibfnamefont {J.}~\bibnamefont
  {Polchinski}},\ }\href@noop {} {\emph {\bibinfo {title} {String {Theory}}}},\
  \bibinfo {series} {Cambridge {Monographs} on {Mathematical} {Physics}}, Vol.\
  \bibinfo {volume} {1–2}\ (\bibinfo  {publisher} {Cambridge University
  Press},\ \bibinfo {address} {Cambridge, UK},\ \bibinfo {year}
  {1998})\BibitemShut {NoStop}%
\bibitem [{\citenamefont {Polyakov}(2018)}]{polyakov_gauge_2018}%
  \BibitemOpen
  \bibfield  {author} {\bibinfo {author} {\bibfnamefont {A.}~\bibnamefont
  {Polyakov}},\ }\href {https://doi.org/10.1201/9780203755082}
  {{\selectlanguage {en}\emph {\bibinfo {title} {Gauge {Fields} and
  {Strings}}}}},\ \bibinfo {edition} {1st}\ ed.\ (\bibinfo  {publisher}
  {Routledge},\ \bibinfo {address} {London},\ \bibinfo {year}
  {2018})\BibitemShut {NoStop}%
\bibitem [{\citenamefont {Klebanov}\ and\ \citenamefont
  {Polyakov}(2002)}]{klebanov_ads_2002}%
  \BibitemOpen
  \bibfield  {author} {\bibinfo {author} {\bibfnamefont {I.}~\bibnamefont
  {Klebanov}}\ and\ \bibinfo {author} {\bibfnamefont {A.}~\bibnamefont
  {Polyakov}},\ }\href {https://doi.org/10.1016/S0370-2693(02)02980-5}
  {\bibfield  {journal} {\bibinfo  {journal} {Physics Letters B}\ }\textbf
  {\bibinfo {volume} {550}},\ \bibinfo {pages} {213} (\bibinfo {year}
  {2002})}\BibitemShut {NoStop}%
\bibitem [{\citenamefont {Guida}\ and\ \citenamefont
  {Zinn-Justin}(1998)}]{guida_critical_1998}%
  \BibitemOpen
  \bibfield  {author} {\bibinfo {author} {\bibfnamefont {R.}~\bibnamefont
  {Guida}}\ and\ \bibinfo {author} {\bibfnamefont {J.}~\bibnamefont
  {Zinn-Justin}},\ }\href {https://doi.org/10.1088/0305-4470/31/40/006}
  {\bibfield  {journal} {\bibinfo  {journal} {Journal of Physics A:
  Mathematical and General}\ }\textbf {\bibinfo {volume} {31}},\ \bibinfo
  {pages} {8103} (\bibinfo {year} {1998})}\BibitemShut {NoStop}%
\bibitem [{\citenamefont {Aarts}\ \emph {et~al.}(2023)\citenamefont {Aarts},
  \citenamefont {Aichelin}, \citenamefont {Allton}, \citenamefont
  {Athenodorou}, \citenamefont {Bachtis}, \citenamefont {Bonanno},
  \citenamefont {Brambilla}, \citenamefont {Bratkovskaya}, \citenamefont
  {Bruno}, \citenamefont {Caselle}, \citenamefont {Conti}, \citenamefont
  {Contino}, \citenamefont {Cosmai}, \citenamefont {Cuteri}, \citenamefont
  {Del~Debbio}, \citenamefont {D’Elia}, \citenamefont {Dimopoulos},
  \citenamefont {Di~Renzo}, \citenamefont {Galatyuk}, \citenamefont {Guenther},
  \citenamefont {Houtz}, \citenamefont {Karsch}, \citenamefont {Kotov},
  \citenamefont {Lombardo}, \citenamefont {Lucini}, \citenamefont {Maio},
  \citenamefont {Panero}, \citenamefont {Pawlowski}, \citenamefont
  {Pelissetto}, \citenamefont {Philipsen}, \citenamefont {Rago}, \citenamefont
  {Ratti}, \citenamefont {Ryan}, \citenamefont {Sannino}, \citenamefont
  {Sasaki}, \citenamefont {Schicho}, \citenamefont {Schmidt}, \citenamefont
  {Sharma}, \citenamefont {Soloveva}, \citenamefont {Sorba},\ and\
  \citenamefont {Wiese}}]{aarts_phase_2023}%
  \BibitemOpen
  \bibfield  {author} {\bibinfo {author} {\bibfnamefont {G.}~\bibnamefont
  {Aarts}}, \bibinfo {author} {\bibfnamefont {J.}~\bibnamefont {Aichelin}},
  \bibinfo {author} {\bibfnamefont {C.}~\bibnamefont {Allton}}, \bibinfo
  {author} {\bibfnamefont {A.}~\bibnamefont {Athenodorou}}, \bibinfo {author}
  {\bibfnamefont {D.}~\bibnamefont {Bachtis}}, \bibinfo {author} {\bibfnamefont
  {C.}~\bibnamefont {Bonanno}}, \bibinfo {author} {\bibfnamefont
  {N.}~\bibnamefont {Brambilla}}, \bibinfo {author} {\bibfnamefont
  {E.}~\bibnamefont {Bratkovskaya}}, \bibinfo {author} {\bibfnamefont
  {M.}~\bibnamefont {Bruno}}, \bibinfo {author} {\bibfnamefont
  {M.}~\bibnamefont {Caselle}}, \bibinfo {author} {\bibfnamefont
  {C.}~\bibnamefont {Conti}}, \bibinfo {author} {\bibfnamefont
  {R.}~\bibnamefont {Contino}}, \bibinfo {author} {\bibfnamefont
  {L.}~\bibnamefont {Cosmai}}, \bibinfo {author} {\bibfnamefont
  {F.}~\bibnamefont {Cuteri}}, \bibinfo {author} {\bibfnamefont
  {L.}~\bibnamefont {Del~Debbio}}, \bibinfo {author} {\bibfnamefont
  {M.}~\bibnamefont {D’Elia}}, \bibinfo {author} {\bibfnamefont
  {P.}~\bibnamefont {Dimopoulos}}, \bibinfo {author} {\bibfnamefont
  {F.}~\bibnamefont {Di~Renzo}}, \bibinfo {author} {\bibfnamefont
  {T.}~\bibnamefont {Galatyuk}}, \bibinfo {author} {\bibfnamefont {J.~N.}\
  \bibnamefont {Guenther}}, \bibinfo {author} {\bibfnamefont {R.}~\bibnamefont
  {Houtz}}, \bibinfo {author} {\bibfnamefont {F.}~\bibnamefont {Karsch}},
  \bibinfo {author} {\bibfnamefont {A.~Y.}\ \bibnamefont {Kotov}}, \bibinfo
  {author} {\bibfnamefont {M.~P.}\ \bibnamefont {Lombardo}}, \bibinfo {author}
  {\bibfnamefont {B.}~\bibnamefont {Lucini}}, \bibinfo {author} {\bibfnamefont
  {L.}~\bibnamefont {Maio}}, \bibinfo {author} {\bibfnamefont {M.}~\bibnamefont
  {Panero}}, \bibinfo {author} {\bibfnamefont {J.~M.}\ \bibnamefont
  {Pawlowski}}, \bibinfo {author} {\bibfnamefont {A.}~\bibnamefont
  {Pelissetto}}, \bibinfo {author} {\bibfnamefont {O.}~\bibnamefont
  {Philipsen}}, \bibinfo {author} {\bibfnamefont {A.}~\bibnamefont {Rago}},
  \bibinfo {author} {\bibfnamefont {C.}~\bibnamefont {Ratti}}, \bibinfo
  {author} {\bibfnamefont {S.~M.}\ \bibnamefont {Ryan}}, \bibinfo {author}
  {\bibfnamefont {F.}~\bibnamefont {Sannino}}, \bibinfo {author} {\bibfnamefont
  {C.}~\bibnamefont {Sasaki}}, \bibinfo {author} {\bibfnamefont
  {P.}~\bibnamefont {Schicho}}, \bibinfo {author} {\bibfnamefont
  {C.}~\bibnamefont {Schmidt}}, \bibinfo {author} {\bibfnamefont
  {S.}~\bibnamefont {Sharma}}, \bibinfo {author} {\bibfnamefont
  {O.}~\bibnamefont {Soloveva}}, \bibinfo {author} {\bibfnamefont
  {M.}~\bibnamefont {Sorba}},\ and\ \bibinfo {author} {\bibfnamefont {U.-J.}\
  \bibnamefont {Wiese}},\ }\href {https://doi.org/10.1016/j.ppnp.2023.104070}
  {\bibfield  {journal} {\bibinfo  {journal} {Progress in Particle and Nuclear
  Physics}\ }\textbf {\bibinfo {volume} {133}},\ \bibinfo {pages} {104070}
  (\bibinfo {year} {2023})}\BibitemShut {NoStop}%
\bibitem [{\citenamefont {Stanley}(1968)}]{stanley_dependence_1968}%
  \BibitemOpen
  \bibfield  {author} {\bibinfo {author} {\bibfnamefont {H.~E.}\ \bibnamefont
  {Stanley}},\ }\href {https://doi.org/10.1103/PhysRevLett.20.589} {\bibfield
  {journal} {\bibinfo  {journal} {Physical Review Letters}\ }\textbf {\bibinfo
  {volume} {20}},\ \bibinfo {pages} {589} (\bibinfo {year} {1968})}\BibitemShut
  {NoStop}%
\bibitem [{\citenamefont {Wilson}\ and\ \citenamefont
  {Fisher}(1972)}]{wilson_critical_1972}%
  \BibitemOpen
  \bibfield  {author} {\bibinfo {author} {\bibfnamefont {K.~G.}\ \bibnamefont
  {Wilson}}\ and\ \bibinfo {author} {\bibfnamefont {M.~E.}\ \bibnamefont
  {Fisher}},\ }\href {https://doi.org/10.1103/PhysRevLett.28.240} {\bibfield
  {journal} {\bibinfo  {journal} {Physical Review Letters}\ }\textbf {\bibinfo
  {volume} {28}},\ \bibinfo {pages} {240} (\bibinfo {year} {1972})}\BibitemShut
  {NoStop}%
\bibitem [{\citenamefont {Parisi}(1988)}]{parisi_statistical_1988}%
  \BibitemOpen
  \bibfield  {author} {\bibinfo {author} {\bibfnamefont {G.}~\bibnamefont
  {Parisi}},\ }\href@noop {} {\emph {\bibinfo {title} {Statistical {Field}
  {Theory}}}},\ \bibinfo {series} {Frontiers in {Physics}}, Vol.~\bibinfo
  {volume} {66}\ (\bibinfo  {publisher} {Addison-Wesley},\ \bibinfo {address}
  {Redwood City, CA},\ \bibinfo {year} {1988})\BibitemShut {NoStop}%
\bibitem [{\citenamefont {Metlitski}(2021)}]{metlitski_boundary_2021}%
  \BibitemOpen
  \bibfield  {author} {\bibinfo {author} {\bibfnamefont {M.~A.}\ \bibnamefont
  {Metlitski}},\ }\href {https://doi.org/10.48550/arXiv.2009.05119} {\bibinfo
  {title} {Boundary criticality of the {O}({N}) model in d = 3 critically
  revisited}} (\bibinfo {year} {2021}),\ \bibinfo {note} {arXiv:2009.05119
  [cond-mat.str-el]}\BibitemShut {NoStop}%
\bibitem [{\citenamefont {Kos}\ \emph {et~al.}(2014)\citenamefont {Kos},
  \citenamefont {Poland},\ and\ \citenamefont
  {Simmons-Duffin}}]{kos_bootstrapping_2014}%
  \BibitemOpen
  \bibfield  {author} {\bibinfo {author} {\bibfnamefont {F.}~\bibnamefont
  {Kos}}, \bibinfo {author} {\bibfnamefont {D.}~\bibnamefont {Poland}},\ and\
  \bibinfo {author} {\bibfnamefont {D.}~\bibnamefont {Simmons-Duffin}},\ }\href
  {https://doi.org/10.1007/JHEP06(2014)091} {\bibfield  {journal} {\bibinfo
  {journal} {Journal of High Energy Physics}\ }\textbf {\bibinfo {volume}
  {2014}},\ \bibinfo {pages} {91} (\bibinfo {year} {2014})}\BibitemShut
  {NoStop}%
\bibitem [{\citenamefont {Schölkopf}\ and\ \citenamefont
  {Smola}(2018)}]{scholkopf_learning_2018}%
  \BibitemOpen
  \bibfield  {author} {\bibinfo {author} {\bibfnamefont {B.}~\bibnamefont
  {Schölkopf}}\ and\ \bibinfo {author} {\bibfnamefont {A.}~\bibnamefont
  {Smola}},\ }\href@noop {} {\emph {\bibinfo {title} {Learning with {Kernels}:
  {Support} {Vector} {Machines}, {Regularization}, {Optimization}, and
  {Beyond}}}}\ (\bibinfo  {publisher} {The MIT Press},\ \bibinfo {year}
  {2018})\BibitemShut {NoStop}%
\bibitem [{\citenamefont {Mikolov}\ \emph {et~al.}(2013)\citenamefont
  {Mikolov}, \citenamefont {Chen}, \citenamefont {Corrado},\ and\ \citenamefont
  {Dean}}]{mikolov_efficient_2013}%
  \BibitemOpen
  \bibfield  {author} {\bibinfo {author} {\bibfnamefont {T.}~\bibnamefont
  {Mikolov}}, \bibinfo {author} {\bibfnamefont {K.}~\bibnamefont {Chen}},
  \bibinfo {author} {\bibfnamefont {G.}~\bibnamefont {Corrado}},\ and\ \bibinfo
  {author} {\bibfnamefont {J.}~\bibnamefont {Dean}},\ }\href
  {https://doi.org/10.48550/arXiv.1301.3781} {\bibinfo {title} {Efficient
  {Estimation} of {Word} {Representations} in {Vector} {Space}}} (\bibinfo
  {year} {2013}),\ \bibinfo {note} {arXiv:1301.3781 [cs.CL]}\BibitemShut
  {NoStop}%
\bibitem [{\citenamefont {Vaswani}\ \emph {et~al.}(2017)\citenamefont
  {Vaswani}, \citenamefont {Shazeer}, \citenamefont {Parmar}, \citenamefont
  {Uszkoreit}, \citenamefont {Jones}, \citenamefont {Gomez}, \citenamefont
  {Kaiser},\ and\ \citenamefont {Polosukhin}}]{vaswani_attention_2017}%
  \BibitemOpen
  \bibfield  {author} {\bibinfo {author} {\bibfnamefont {A.}~\bibnamefont
  {Vaswani}}, \bibinfo {author} {\bibfnamefont {N.}~\bibnamefont {Shazeer}},
  \bibinfo {author} {\bibfnamefont {N.}~\bibnamefont {Parmar}}, \bibinfo
  {author} {\bibfnamefont {J.}~\bibnamefont {Uszkoreit}}, \bibinfo {author}
  {\bibfnamefont {L.}~\bibnamefont {Jones}}, \bibinfo {author} {\bibfnamefont
  {A.~N.}\ \bibnamefont {Gomez}}, \bibinfo {author} {\bibfnamefont
  {L.}~\bibnamefont {Kaiser}},\ and\ \bibinfo {author} {\bibfnamefont
  {I.}~\bibnamefont {Polosukhin}},\ }in\ \href@noop {} {\emph {\bibinfo
  {booktitle} {Advances in {Neural} {Information} {Processing} {Systems}}}},\
  Vol.~\bibinfo {volume} {30}\ (\bibinfo {year} {2017})\BibitemShut {NoStop}%
\bibitem [{\citenamefont {Calvanese~Strinati}\ \emph
  {et~al.}(2021)\citenamefont {Calvanese~Strinati}, \citenamefont {Bello},
  \citenamefont {Dalla~Torre},\ and\ \citenamefont
  {Pe'er}}]{PhysRevLett.126.143901}%
  \BibitemOpen
  \bibfield  {author} {\bibinfo {author} {\bibfnamefont {M.}~\bibnamefont
  {Calvanese~Strinati}}, \bibinfo {author} {\bibfnamefont {L.}~\bibnamefont
  {Bello}}, \bibinfo {author} {\bibfnamefont {E.~G.}\ \bibnamefont
  {Dalla~Torre}},\ and\ \bibinfo {author} {\bibfnamefont {A.}~\bibnamefont
  {Pe'er}},\ }\href {https://doi.org/10.1103/PhysRevLett.126.143901} {\bibfield
   {journal} {\bibinfo  {journal} {Phys. Rev. Lett.}\ }\textbf {\bibinfo
  {volume} {126}},\ \bibinfo {pages} {143901} (\bibinfo {year}
  {2021})}\BibitemShut {NoStop}%
\bibitem [{\citenamefont {Calvanese~Strinati}\ and\ \citenamefont
  {Conti}(2022)}]{calvanese_strinati_multidimensional_2022}%
  \BibitemOpen
  \bibfield  {author} {\bibinfo {author} {\bibfnamefont {M.}~\bibnamefont
  {Calvanese~Strinati}}\ and\ \bibinfo {author} {\bibfnamefont
  {C.}~\bibnamefont {Conti}},\ }\href
  {https://doi.org/10.1038/s41467-022-34847-9} {\bibfield  {journal} {\bibinfo
  {journal} {Nature Communications}\ }\textbf {\bibinfo {volume} {13}},\
  \bibinfo {pages} {7248} (\bibinfo {year} {2022})}\BibitemShut {NoStop}%
\bibitem [{\citenamefont {Ben-Ami}\ \emph {et~al.}(2023)\citenamefont
  {Ben-Ami}, \citenamefont {Aharonovich},\ and\ \citenamefont
  {Pe'er}}]{Ben-Ami:23}%
  \BibitemOpen
  \bibfield  {author} {\bibinfo {author} {\bibfnamefont {S.}~\bibnamefont
  {Ben-Ami}}, \bibinfo {author} {\bibfnamefont {I.}~\bibnamefont
  {Aharonovich}},\ and\ \bibinfo {author} {\bibfnamefont {A.}~\bibnamefont
  {Pe'er}},\ }\href {https://doi.org/10.1364/OE.482828} {\bibfield  {journal}
  {\bibinfo  {journal} {Opt. Express}\ }\textbf {\bibinfo {volume} {31}},\
  \bibinfo {pages} {9264} (\bibinfo {year} {2023})}\BibitemShut {NoStop}%
\bibitem [{\citenamefont {Strinati}\ \emph {et~al.}(2026)\citenamefont
  {Strinati}, \citenamefont {Gentilini}, \citenamefont {Ghofraniha},
  \citenamefont {Pierangeli},\ and\ \citenamefont
  {Conti}}]{strinati_optical_2026}%
  \BibitemOpen
  \bibfield  {author} {\bibinfo {author} {\bibfnamefont {M.~C.}\ \bibnamefont
  {Strinati}}, \bibinfo {author} {\bibfnamefont {S.}~\bibnamefont {Gentilini}},
  \bibinfo {author} {\bibfnamefont {N.}~\bibnamefont {Ghofraniha}}, \bibinfo
  {author} {\bibfnamefont {D.}~\bibnamefont {Pierangeli}},\ and\ \bibinfo
  {author} {\bibfnamefont {C.}~\bibnamefont {Conti}},\ }\href
  {https://doi.org/10.1364/AOP.574825} {\bibfield  {journal} {\bibinfo
  {journal} {Advances in Optics and Photonics}\ }\textbf {\bibinfo {volume}
  {18}},\ \bibinfo {pages} {421} (\bibinfo {year} {2026})}\BibitemShut
  {NoStop}%
\bibitem [{\citenamefont {Cummins}\ and\ \citenamefont
  {Berloff}(2025)}]{cummins_vector_2025}%
  \BibitemOpen
  \bibfield  {author} {\bibinfo {author} {\bibfnamefont {J.~S.}\ \bibnamefont
  {Cummins}}\ and\ \bibinfo {author} {\bibfnamefont {N.~G.}\ \bibnamefont
  {Berloff}},\ }\href {https://doi.org/10.1038/s42005-025-02145-7} {\bibfield
  {journal} {\bibinfo  {journal} {Communications Physics}\ }\textbf {\bibinfo
  {volume} {8}},\ \bibinfo {pages} {225} (\bibinfo {year} {2025})}\BibitemShut
  {NoStop}%
\bibitem [{\citenamefont {Syed}\ \emph {et~al.}(2026)\citenamefont {Syed},
  \citenamefont {Wang},\ and\ \citenamefont {Berloff}}]{syed_soft_2026}%
  \BibitemOpen
  \bibfield  {author} {\bibinfo {author} {\bibfnamefont {M.}~\bibnamefont
  {Syed}}, \bibinfo {author} {\bibfnamefont {R.~Z.}\ \bibnamefont {Wang}},\
  and\ \bibinfo {author} {\bibfnamefont {N.~G.}\ \bibnamefont {Berloff}},\
  }\href {https://doi.org/10.48550/ARXIV.2604.01003} {\bibinfo {title} {Soft
  vector spins with dimensional annealing for combinatorial optimization}}
  (\bibinfo {year} {2026}),\ \bibinfo {note} {version Number: 1}\BibitemShut
  {NoStop}%
\bibitem [{\citenamefont {Calvanese~Strinati}\ and\ \citenamefont
  {Conti}(2024)}]{calvanese_strinati_hyperscaling_2024}%
  \BibitemOpen
  \bibfield  {author} {\bibinfo {author} {\bibfnamefont {M.}~\bibnamefont
  {Calvanese~Strinati}}\ and\ \bibinfo {author} {\bibfnamefont
  {C.}~\bibnamefont {Conti}},\ }\href
  {https://doi.org/10.1103/PhysRevLett.132.017301} {\bibfield  {journal}
  {\bibinfo  {journal} {Physical Review Letters}\ }\textbf {\bibinfo {volume}
  {132}},\ \bibinfo {pages} {017301} (\bibinfo {year} {2024})}\BibitemShut
  {NoStop}%
\bibitem [{\citenamefont {Calvanese~Strinati}\ and\ \citenamefont
  {Conti}(2025)}]{calvanese_strinati_equalized_2025}%
  \BibitemOpen
  \bibfield  {author} {\bibinfo {author} {\bibfnamefont {M.}~\bibnamefont
  {Calvanese~Strinati}}\ and\ \bibinfo {author} {\bibfnamefont
  {C.}~\bibnamefont {Conti}},\ }\href {https://doi.org/10.1103/cxjt-53qy}
  {\bibfield  {journal} {\bibinfo  {journal} {Physical Review A}\ }\textbf
  {\bibinfo {volume} {112}},\ \bibinfo {pages} {053505} (\bibinfo {year}
  {2025})}\BibitemShut {NoStop}%
\bibitem [{\citenamefont {Chiavazzo}\ \emph {et~al.}(2025)\citenamefont
  {Chiavazzo}, \citenamefont {Calvanese~Strinati}, \citenamefont {Conti},\ and\
  \citenamefont {Pierangeli}}]{chiavazzo_ising_2025}%
  \BibitemOpen
  \bibfield  {author} {\bibinfo {author} {\bibfnamefont {S.}~\bibnamefont
  {Chiavazzo}}, \bibinfo {author} {\bibfnamefont {M.}~\bibnamefont
  {Calvanese~Strinati}}, \bibinfo {author} {\bibfnamefont {C.}~\bibnamefont
  {Conti}},\ and\ \bibinfo {author} {\bibfnamefont {D.}~\bibnamefont
  {Pierangeli}},\ }\href {https://doi.org/10.1103/qs29-2xqc} {\bibfield
  {journal} {\bibinfo  {journal} {Physical Review Letters}\ }\textbf {\bibinfo
  {volume} {135}},\ \bibinfo {pages} {063801} (\bibinfo {year}
  {2025})}\BibitemShut {NoStop}%
\bibitem [{\citenamefont {Kalnins}\ \emph {et~al.}(2002)\citenamefont
  {Kalnins}, \citenamefont {Miller},\ and\ \citenamefont
  {Pogosyan}}]{kalnins2002ndimensional}%
  \BibitemOpen
  \bibfield  {author} {\bibinfo {author} {\bibfnamefont {E.~G.}\ \bibnamefont
  {Kalnins}}, \bibinfo {author} {\bibfnamefont {W.}~\bibnamefont {Miller}},\
  and\ \bibinfo {author} {\bibfnamefont {G.~S.}\ \bibnamefont {Pogosyan}},\
  }\href {https://doi.org/10.1134/1.1490116} {\bibfield  {journal} {\bibinfo
  {journal} {Phys. Atom. Nuclei}\ }\textbf {\bibinfo {volume} {65}},\ \bibinfo
  {pages} {1086} (\bibinfo {year} {2002})}\BibitemShut {NoStop}%
\bibitem [{\citenamefont {Cohl}(2011)}]{cohl2011ndimensional}%
  \BibitemOpen
  \bibfield  {author} {\bibinfo {author} {\bibfnamefont {H.~S.}\ \bibnamefont
  {Cohl}},\ }\href {https://doi.org/10.3842/SIGMA.2011.108} {\bibfield
  {journal} {\bibinfo  {journal} {SIGMA}\ }\textbf {\bibinfo {volume} {7}},\
  \bibinfo {pages} {108} (\bibinfo {year} {2011})}\BibitemShut {NoStop}%
\bibitem [{\citenamefont {Yamashita}\ \emph {et~al.}(2023)\citenamefont
  {Yamashita}, \citenamefont {Okubo}, \citenamefont {Shimomura}, \citenamefont
  {Ogura}, \citenamefont {Tanida},\ and\ \citenamefont
  {Suzuki}}]{yamashita_low-rank_2023}%
  \BibitemOpen
  \bibfield  {author} {\bibinfo {author} {\bibfnamefont {H.}~\bibnamefont
  {Yamashita}}, \bibinfo {author} {\bibfnamefont {K.-i.}\ \bibnamefont
  {Okubo}}, \bibinfo {author} {\bibfnamefont {S.}~\bibnamefont {Shimomura}},
  \bibinfo {author} {\bibfnamefont {Y.}~\bibnamefont {Ogura}}, \bibinfo
  {author} {\bibfnamefont {J.}~\bibnamefont {Tanida}},\ and\ \bibinfo {author}
  {\bibfnamefont {H.}~\bibnamefont {Suzuki}},\ }\href
  {https://doi.org/10.1103/PhysRevLett.131.063801} {\bibfield  {journal}
  {\bibinfo  {journal} {Physical Review Letters}\ }\textbf {\bibinfo {volume}
  {131}},\ \bibinfo {pages} {063801} (\bibinfo {year} {2023})}\BibitemShut
  {NoStop}%
\bibitem [{\citenamefont {Veraldi}\ \emph {et~al.}(2025)\citenamefont
  {Veraldi}, \citenamefont {Pierangeli}, \citenamefont {Gentilini},
  \citenamefont {Strinati}, \citenamefont {Sakellariou}, \citenamefont
  {Cummins}, \citenamefont {Kamaletdinov}, \citenamefont {Syed}, \citenamefont
  {Wang}, \citenamefont {Berloff}, \citenamefont {Karanikolopoulos},
  \citenamefont {Savvidis},\ and\ \citenamefont {Conti}}]{veraldi_fully_2025}%
  \BibitemOpen
  \bibfield  {author} {\bibinfo {author} {\bibfnamefont {D.}~\bibnamefont
  {Veraldi}}, \bibinfo {author} {\bibfnamefont {D.}~\bibnamefont {Pierangeli}},
  \bibinfo {author} {\bibfnamefont {S.}~\bibnamefont {Gentilini}}, \bibinfo
  {author} {\bibfnamefont {M.~C.}\ \bibnamefont {Strinati}}, \bibinfo {author}
  {\bibfnamefont {J.}~\bibnamefont {Sakellariou}}, \bibinfo {author}
  {\bibfnamefont {J.~S.}\ \bibnamefont {Cummins}}, \bibinfo {author}
  {\bibfnamefont {A.}~\bibnamefont {Kamaletdinov}}, \bibinfo {author}
  {\bibfnamefont {M.}~\bibnamefont {Syed}}, \bibinfo {author} {\bibfnamefont
  {R.~Z.}\ \bibnamefont {Wang}}, \bibinfo {author} {\bibfnamefont {N.~G.}\
  \bibnamefont {Berloff}}, \bibinfo {author} {\bibfnamefont {D.}~\bibnamefont
  {Karanikolopoulos}}, \bibinfo {author} {\bibfnamefont {P.~G.}\ \bibnamefont
  {Savvidis}},\ and\ \bibinfo {author} {\bibfnamefont {C.}~\bibnamefont
  {Conti}},\ }\href {https://doi.org/10.1103/PhysRevLett.134.063802} {\bibfield
   {journal} {\bibinfo  {journal} {Physical Review Letters}\ }\textbf {\bibinfo
  {volume} {134}},\ \bibinfo {pages} {063802} (\bibinfo {year}
  {2025})}\BibitemShut {NoStop}%
\bibitem [{\citenamefont {Mendoza-Yero}\ \emph {et~al.}(2014)\citenamefont
  {Mendoza-Yero}, \citenamefont {Mínguez-Vega},\ and\ \citenamefont
  {Lancis}}]{mendoza-yero_encoding_2014}%
  \BibitemOpen
  \bibfield  {author} {\bibinfo {author} {\bibfnamefont {O.}~\bibnamefont
  {Mendoza-Yero}}, \bibinfo {author} {\bibfnamefont {G.}~\bibnamefont
  {Mínguez-Vega}},\ and\ \bibinfo {author} {\bibfnamefont {J.}~\bibnamefont
  {Lancis}},\ }\href {https://doi.org/10.1364/OL.39.001740} {\bibfield
  {journal} {\bibinfo  {journal} {Optics Letters}\ }\textbf {\bibinfo {volume}
  {39}},\ \bibinfo {pages} {1740} (\bibinfo {year} {2014})}\BibitemShut
  {NoStop}%
\bibitem [{\citenamefont {Luo}\ \emph {et~al.}(2023)\citenamefont {Luo},
  \citenamefont {Mi}, \citenamefont {Huang},\ and\ \citenamefont
  {Ruan}}]{luo_wavelength-division_2023}%
  \BibitemOpen
  \bibfield  {author} {\bibinfo {author} {\bibfnamefont {L.}~\bibnamefont
  {Luo}}, \bibinfo {author} {\bibfnamefont {Z.}~\bibnamefont {Mi}}, \bibinfo
  {author} {\bibfnamefont {J.}~\bibnamefont {Huang}},\ and\ \bibinfo {author}
  {\bibfnamefont {Z.}~\bibnamefont {Ruan}},\ }\href
  {https://doi.org/10.1126/sciadv.adg6238} {\bibfield  {journal} {\bibinfo
  {journal} {Science Advances}\ }\textbf {\bibinfo {volume} {9}},\ \bibinfo
  {pages} {eadg6238} (\bibinfo {year} {2023})}\BibitemShut {NoStop}%
\bibitem [{\citenamefont {Mohseni}\ \emph {et~al.}(2022)\citenamefont
  {Mohseni}, \citenamefont {McMahon},\ and\ \citenamefont
  {Byrnes}}]{mohseni_ising_2022}%
  \BibitemOpen
  \bibfield  {author} {\bibinfo {author} {\bibfnamefont {N.}~\bibnamefont
  {Mohseni}}, \bibinfo {author} {\bibfnamefont {P.~L.}\ \bibnamefont
  {McMahon}},\ and\ \bibinfo {author} {\bibfnamefont {T.}~\bibnamefont
  {Byrnes}},\ }\href {https://doi.org/10.1038/s42254-022-00440-8} {\bibfield
  {journal} {\bibinfo  {journal} {Nature Reviews Physics}\ }\textbf {\bibinfo
  {volume} {4}},\ \bibinfo {pages} {363} (\bibinfo {year} {2022})}\BibitemShut
  {NoStop}%
\bibitem [{\citenamefont {Davis}(1994)}]{davis1994circulant}%
  \BibitemOpen
  \bibfield  {author} {\bibinfo {author} {\bibfnamefont {P.~J.}\ \bibnamefont
  {Davis}},\ }\href@noop {} {\emph {\bibinfo {title} {Circulant Matrices}}},\
  Chelsea Publishing Series\ (\bibinfo  {publisher} {Chelsea},\ \bibinfo {year}
  {1994})\BibitemShut {NoStop}%
\end{thebibliography}%

\end{document}